# Transient Characteristics of β-Ga$_2$O$_3$ Nanomembrane Schottky Barrier Diodes on Various Substrates

Junyu Lai[1], Jung-Hun Seo[1, *]

[1] Department of Materials Design and Innovation, University at Buffalo, The State University of New York, Buffalo, NY USA 14260.

E-mail: junghuns@buffalo.edu

## Abstract

In this paper, a transient delayed rising and fall time of β-Ga$_2$O$_3$ NMs Schottky barrier diodes (SBDs) formed on four different substrates (diamond, Si, sapphire, and polyimide) were measured using a sub-micron second resolution time-resolved electrical measurement system under a different temperature condition. The devices exhibited noticeably less-delayed turn-on-/off- the transient time when β- Ga$_2$O$_3$ NMs SBDs were transfer-printed on a high-k substrate. Furthermore, a relationship between the β- Ga$_2$O$_3$ NM thicknesses and their transient characteristics were systematically investigated and found that phonon scattering plays an important role in heat dissipation as the thickness of β- Ga$_2$O$_3$ NMs get thinner which is also verified by the Multiphysics simulator. Overall, our result reveals the impact of various substrates with different thermal properties and different β- Ga$_2$O$_3$ NMs thickness with the performance of β- Ga$_2$O$_3$ NMs based devices. Hence, these results can guide further efforts us to optimize the performance of future β- Ga$_2$O$_3$ devices by maximizing heat dissipation from the β-Ga2O3 layer.

## 1. Introduction

Beta gallium oxide (β-Ga$_2$O$_3$) has attracted considerable attention as a next-generation wide-bandgap semiconductor for power device applications, high-frequency devices, and solar-blind photodetectors due to its superior material properties [1, 2], such as an ultrawide bandgap value (>4.8 eV), good carrier mobility (>300 cm$^2$v$^{-1}$s$^{-1}$), and critical breakdown electric field (≈ 8 MV cm$^{-1}$)[3, 4]. To date, several β-Ga$_2$O$_3$ based field-effect transistors, gigahertz range RF amplifiers, and solar-blind Schottky diodes have been successfully demonstrated [5-7]. However, one critical native disadvantage of β-Ga$_2$O$_3$ is its extremely low thermal conductivity (*k*) (10~ 25 W/m·K) [8, 9], even compared with other popular wide bandgap semiconductors such as silicon carbide (SiC) at 387 W/m·K [10], aluminum nitride (AlN) at 200 W/m·K [11], and gallium nitride (GaN) at 170 W/m·K [12]. Because the applications above generate a significant amount of Joule heat during operation [13-15], a poor thermal property of β-Ga$_2$O$_3$ would become more pronounced in β-Ga$_2$O$_3$ based devices, which causes an even more increase in temperature and a nonuniform distribution of dissipated power, thus emerging as one of the most serious concerns in the degradation of variability and reliability in β-Ga$_2$O$_3$ based applications. [16-18]

Since thermal management is critically important for the efficient operation of β-Ga$_2$O$_3$ based devices, several studies have been carried out to deal with the heat dissipation issue in β-Ga$_2$O$_3$. For example, Rimmon et al. [19] and A.O Nell et al. [20] reported a method of reducing the self-heating effect by straining silicon substrate and target material, leading to an increase in carrier mobility and thus balancing the effect of self-heat. However, this technology suffers two main challenges: it is difficult to control the defect densities between the interface of Si and target materials to control Si's strain relaxation levels. Another method by Li et al. [21, 22] and Shi et al. [21, 22] proposed a method of mitigating the low thermal conductivity issue, increasing the thermal boundary conductance, and lowering the thermal resistance of β-Ga$_2$O$_3$ with the assistance of metallic or dielectric interfacial layers. Recently, another novel route to manage a thermal property of β-Ga$_2$O$_3$ was reported by utilizing a sub-micron thin and freestanding format of β-Ga$_2$O$_3$, also called β-Ga$_2$O$_3$ nanomembrane (NM). β-Ga$_2$O$_3$ NMs can be transfer-printed onto any desired substrates. [23-26] Thus, efficient heat dissipation from β-Ga$_2$O$_3$ is expected when β-Ga$_2$O$_3$ NMs are transfer-printed on a high-k substrate. For example, Zheng et al. and Cheng et al. used single-crystal diamond substrate as a heat dissipator to create the β-Ga$_2$O$_3$ NM/diamond structure and successfully characterized the thermal conductivity of β-Ga$_2$O$_3$ NM and thermal boundary conductance at the β-Ga$_2$O$_3$ NM/diamond interface [8, 27, 28]. Cheng et al. and Yixiong et al. also formed β-Ga$_2$O$_3$ NM/SiC structure to realize a similar cooling effect. [29, 30] While all these studies on thermal management strategies on β-Ga$_2$O$_3$ provide insightful information about maximizing material property of β-Ga$_2$O$_3$, a comprehensive study on the impact of the electrical property on the heating effect depending on various substrates has not yet been experimentally demonstrated.

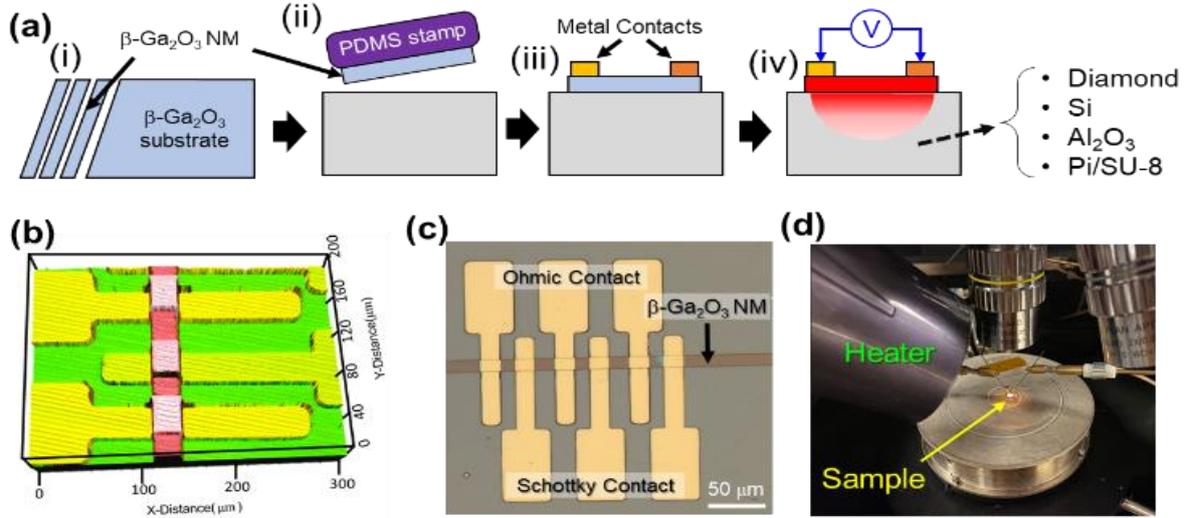

**Figure 1.** (a) A schematic illustration of the β-Ga$_2$O$_3$ NMs SBD fabrication process: (i) a creation of β-Ga$_2$O$_3$ NMs, (ii) a micro-transfer printing of β-Ga$_2$O$_3$ NMs onto four different substrates, (iii) a metallization on β-Ga$_2$O$_3$ NMs to form SBDs, (iv) a pulsed I-V characterization. (b) Three-dimensional morphology and (c) a microscopic image of the fabricated β-Ga$_2$O$_3$ NMs SBDs (d) an image of the heating setup.

In this paper, β-Ga$_2$O$_3$ NMs based Schottky barrier diodes (SBDs) were formed on four different substrates via a micro-transfer printing that has a wide range of thermal conductivity from 0.25 W/m·K to 2200 W/m·K, which include polyimide, sapphire, silicon, and diamond substrate, and characterized a transport property of β-Ga$_2$O$_3$ NMs SBDs under different temperature conditions using a sub-micro second resolution time-resolved electrical measurement system. The devices exhibited noticeably less-delayed turn on/off transient time when β-Ga$_2$O$_3$ NMs SBDs were transfer-printed on a high-k substrate. Furthermore, a relationship between the β-Ga$_2$O$_3$ NM thicknesses and their time-resolved electrical properties was systematically investigated. We found that phonon scattering plays an important role in heat dissipation as the thickness of β-Ga$_2$O$_3$ NMs gets thinner, which is also verified by the Multiphysics simulator. Overall, our result reveals the impact of various substrates with different thermal properties and different β-Ga$_2$O$_3$ NMs thickness with the performance of β-Ga$_2$O$_3$ NMs based devices. Hence, these results can guide future efforts to optimize the performance of future β-Ga$_2$O$_3$ devices by maximizing heat dissipation from the β-Ga$_2$O$_3$ layer.

## 2. Results and discussion

Figure 1(a) shows the schematic illustration of the device fabrication steps. The detail of the fabrication process can be found in the Experimental Section and Ref. [23-25, 31]. Briefly, the fabrication began with a β-Ga$^2$O$^3$ bulk substrate grown by molecular beam epitaxy and moderately Sn doped (concentration of $1\times10^{18}$ cm$^{-3}$). β-Ga$^2$O$^3$ NMs were created by a mechanical exfoliation method (Figure 1(a)(i)). Once β-Ga$^2$O$^3$ NMs were created, we carefully transfer-printed β-Ga$^2$O$^3$ NMs onto four different substrates using an elastomeric stamp (polydimethylsiloxane, PDMS); namely, undoped Si, sapphire, diamond, and polyimide (PI) substrates (Figure 1(a)(ii)). After completing the transfer process, an Ohmic metal stack and a Schottky metal stack were deposited to make Schottky barrier diodes (SBDs) (Figure 1(a)(iii)). Prior to Ohmic metal deposition, a plasma treatment was carried out on β-Ga$^2$O$^3$ NMs by a BCl3/Ar plasma treatment using a reactive ion etcher (RIE) to achieve ohmic contact and avoid an additional high-temperature annealing process. Then, SBDs were measured using a Keithley 4200 SCS semiconductor parameter analyzer with a pulsed I-V unit in a dark box (Figure 1(a)(iv)). Figure 1(b) and (c) show the 3D profile and microscopic images of the as fabricate β-Ga$_2$O$_3$ NMs SBDs on a diamond substrate to show a generic shape of β-Ga$_2$O$_3$ NMs SBD. To compare transient time in β-Ga$_2$O$_3$ NMs SBDs between room temperature and elevated temperature, the devices were exposed in heated air and incandescent lamp with the objective lens for 10 min to heat the top surface of the device, while the stainless-steel probe station stage remains at 20 °C as shown in Figure (d). Figure S1 shows I-V curves measured from β-Ga$_2$O$_3$ NMs SBDs on PI, Si, sapphire, and diamond substrates with different β-Ga$_2$O$_3$ NM thicknesses varying from 300 nm to 1000 nm. All β-Ga$_2$O$_3$ NMs SBDs exhibited a good rectifying behavior with an on/off ratio higher than $10^{-6}$ and the ideality factor (*n*) of 1.8 calculated based on thermionic emission theory [32]. The similarly good current-voltage (I-V) characteristics in β-Ga$_2$O$_3$ NMs SBDs set a baseline for the time-resolved electrical characterization that relies on different substrates.

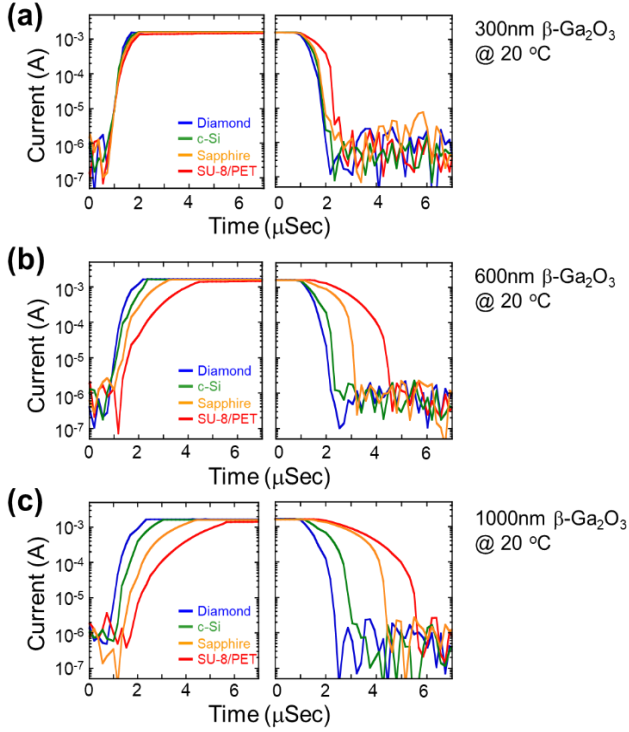

**Figure 2.** A set of pulsed I-V curves of β-Ga$_2$O$_3$ NMs SBDs on four different substrates measured at room temperature with three different β-Ga$_2$O$_3$ NMs thicknesses **(a)** 300 nm, **(b)** 600 nm, **(c)** 1000 nm, respectively.

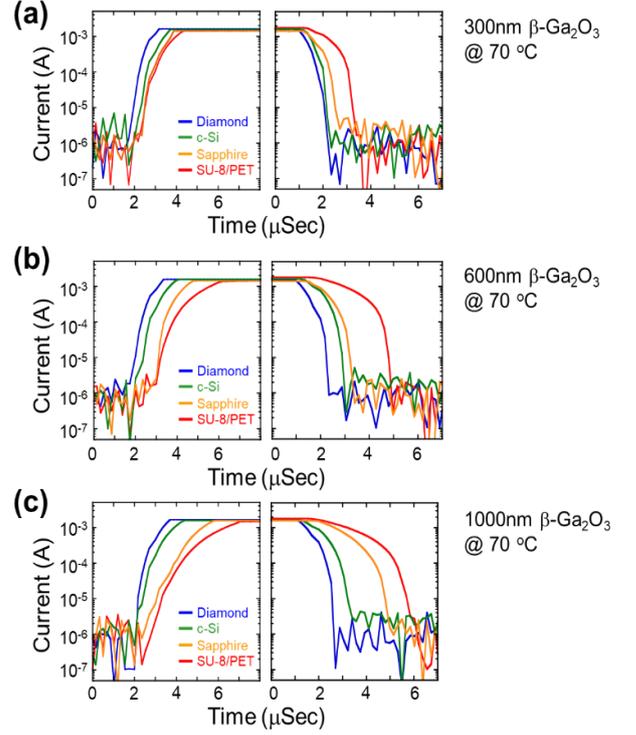

**Figure 3.** A set of pulsed I-V curves of β-Ga$_2$O$_3$ NMs SBDs on four different substrates measured at 70 °C with three different Ga$_2$O$_3$ NMs thicknesses **(a)** 300 nm, **(b)** 600 nm, **(c)** 1000 nm, respectively.

Compared with a typical direct current (DC) measurement, an investigation of transient time measurement using a pulsed I-V provides the isothermal condition of the device, thus offering a more accurate power dissipation effect and thermal resistance characteristics. Figure 2 shows the time-resolved pulsed I-V characteristics of the β-Ga$_2$O$_3$ NMs SBDs on PI, Si, sapphire, and diamond substrates at room temperature. The devices were biased for 1 mSec with a current compliance of 3 mA to prevent unwanted thermal damage. Details about the measurement setup can be found in the Method section. The left panel of Figure 2(a)-(c) are the zoom-in plots of transient rise times on different substrates and different β-Ga$_2$O$_3$ NM thicknesses (from 300 nm to 1000 nm). Similarly, the right panel of Figure 2(a)-(c) are the zoom-in plots of transient fall times on different substrates and the same β-Ga$_2$O$_3$ NM thicknesses, respectively. As shown in Figure 2, the rising and fall time of the devices are noticeably different depending on the substrate. β-Ga$_2$O$_3$ NM SBDs on diamond substrate show the shortest transient time, followed by the device with Si, sapphire, and PI substrate. The rising time for 300 nm thick β-Ga$_2$O$_3$ NMs SBDs on the diamond, Si, sapphire, and PI substrates are measured to be 1.07 µs, 1.35 µs, 1.4 µs, and 2.65 µs, and their respective falling times are 1.07 µs, 1.5 µs, 1.65 µs, 2.85 µs, respectively. Interestingly, it was found that the falling time tends to be longer as β-Ga$_2$O$_3$ NMs SBDs are built on the low-k substrate. The rising and fall time of β-Ga$_2$O$_3$ NMs SBDs on diamond remained almost the same as 1.07 µs, while the fall time for devices on sapphire and PI substrate are increased than their rising from 1.4 µs and 2.65 µs to 1.65 µs and 2.85 µs, respectively. The trapping and de-trapping process in β-Ga$_2$O$_3$ NMs and the interaction between β-Ga$_2$O$_3$ NM and substrate can explain this increase in the fall delay characteristic. The trapping and de-trapping process in semiconductors refers to the electron recombining and releasing process to defects or vacancies. Considering that β-$Ga_2O_3$ has a high level of oxygen vacancies, the trapping and de-trapping process becomes one of the major sources of heat generation in β-$Ga_2O_3$ NMs. When β-$Ga_2O_3$ NM SBDs were biased, a Joule heat would be generated from the top surface of β-$Ga_2O_3$ NMs and begin to accumulate and propagate inside the β-$Ga_2O_3$ NM which is called a "self-heating" effect. This process supplies sufficient activation energy to overcome the trap potential barrier and escape from the trap by the equation (1) [33]:

$$J = n\frac{1}{\tau}\exp\left(-\frac{E_i}{kT}\right) \quad (1)$$

Also, the relationship between the trapping time and rising time can be determined by equation (2) [34]:

$$I = 1 - \exp\left(\frac{-t}{\tau}\right)^{\beta} \quad (2)$$

where J is the current density, $\tau$ is the trapping / de-trapping time constant, t is the trapping /de-trapping period, $E_i$ is the trap energy, $k$ is the Boltzmann constant, T is the temperature, $I$ is the device on current, and $\beta$ is the fitting parameter. When combining equations (1) and (2) with the same amount of the on-current, the traps with a shorter trapping time constant require shorter times to be filled. Due to the self-heating effect, as temperature increases, the trapping time constant increases and enables these traps to take longer time to be filled during the charging process. When devices were heated at the same rate and the input power to the device was the same, the rising time for each case did not vary noticeably. However, when the current is stopped, the heat must be dissipated both to air and substrate. The thermal conductivity values of diamond, Si, and sapphire are higher to 2200, 144, and 40 W/m·K, respectively than that of β-$Ga_2O_3$ NM, and the thermal conductivity of PI is substantially lower to 0.25 W/m·K than that of β-$Ga_2O_3$ NMs. [35-38] Thus, their cooling rates are different. As a result, we conclude that the switching characteristic becomes enhanced when β-$Ga_2O_3$ NMs SBDs are formed on a high-k substrate. This trend is consistent regardless of the thickness of β-$Ga_2O_3$ NMs (both 300 nm thick and 1000 nm thick β-$Ga_2O_3$ NMs), as shown in Figure 2.

To further investigate the relationship between delay time, devices substrates, and β-$Ga_2O_3$ NMs thickness, the same time-resolved pulsed I-V characterizations were performed at the elevated temperature. β-$Ga_2O_3$ NM SBDs were heated from two heating sources (namely, (1) a focused incandescent light bulb via an objective lens, (2) a heat gun) from the top of the device to heat the top surface of the device, not the bottom side of the substrate from the probe station stage because the typical stage heating method in the probe station heats the underneath of substrate, which affects devices differently depending on thermal conductivity of the substrate. Figure 3 shows the time-resolved pulsed I-V characteristics of the β-$Ga_2O_3$ NMs SBDs on PI, Si, sapphire, and diamond substrates at elevated temperatures. The thermal imager was used to measure the surface temperature of β-$Ga_2O_3$ NMs SBDs. The devices were heated until the device temperature reaches 70 ºC prior to the measurement and then biased at 3 mA for 1 mSec. In the left panel of Figure 3(a)-(c), the zoomed-in plots show the transient rise times on four different substrates and different β-$Ga_2O_3$ NM thicknesses. Similarly, the right panel of Figure 3(a)-(c) are the zoomed-in plots of transient fall times on four different substrates and different β-$Ga_2O_3$ NM thicknesses, respectively. The rising time for 300 nm thick β-$Ga_2O_3$ NMs SBDs on diamond, Si, sapphire, and PI substrates were 1.45 µs, 1.55 µs, 1.65 µs, 3.4 µs and their respective falling times are 1.45 µs, 1.6 µs, 1.8 µs, and 3.8 µs, respectively. Similar to the test result at room temperature, the rising and fall time of β-$Ga_2O_3$ NMs SBDs on diamond remained almost the same to be 1.45 µs, while the fall time for devices on PI substrate was increased than their rising time

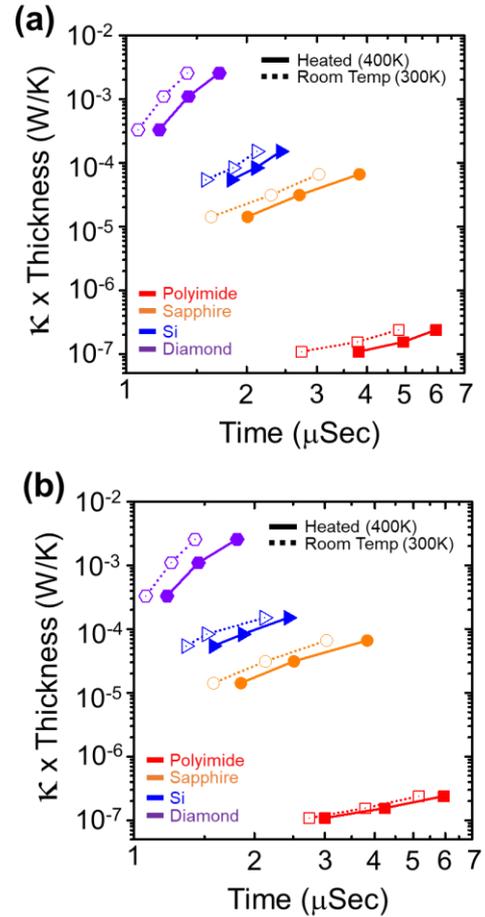

**Figure 4.** A summary of transient delay times of β-$Ga_2O_3$ NMs SBDs under four different substrates as a function of a product of thermal conductivity of the substrate and thickness of β-$Ga_2O_3$ NMs, **(a)** a rising delay times, **(b)** a fall delay times.

from 3.4 µs to 3.8 µs, respectively.

Figure 4(a) and (b) summarize the rising and fall time for β-Ga$_2$O$_3$ NM SBDs, respectively, which are extracted from Figure 2 and Figure 3. To provide a clear view of the relationship among the electrical performance of β-Ga$_2$O$_3$ NM SBDs with respect to the β-Ga$_2$O$_3$ NM thickness and thermal conductivity of the substrate, a transient delay time is plotted with the product of the β-Ga$_2$O$_3$ NM thickness and thermal conductivity of substrate (W/K·Sec: W/K vs. delay time). Therefore, the slope of the curve indicates the capacity of heat dissipation as well as the rate of the trapping/de-trapping process. For the rising time of β-Ga$_2$O$_3$ NM SBDs on the diamond substrate, the slope value is 6.32 ×10$^{-3}$ which is 35, 168, and 1.02×10$^4$ times smaller than the one on Si, sapphire, and PI substrate, respectively. Similar trends were observed in the fall time of β-Ga$_2$O$_3$ NM SBDs on a diamond substrate compared with β-Ga$_2$O$_3$ NM SBDs on other substrates.

To further understand the heat dissipation effect on β-Ga$_2$O$_3$ NMs SBDs, we performed a Multiphysics simulation using the thermal module in COMSOL. The model structure was established using the same dimensional parameters and references [8, 35-39]. In the simulation, the dissipated power was calculated as P = V$_{on}$ ×I$_{on}$, where I$_{on}$ is on state current and V$_{on}$ is the corresponding bias. The actual biasing point was 3 mA at 4 V for all devices, thus an input power density of 12 mW was added to the electrode of the simulated structure. The initial temperature of the entire structure was set to 293.15 K as a boundary condition. Figure 5 and Figure S2 present the heat distribution of β-Ga$_2$O$_3$ NMs with 100 nm, 500 nm, and 1000 nm thicknesses under various substrates at the equilibrium state and the transit state at 1 μSec. As shown in Figure 5, the local temperature at the electrode/β-Ga$_2$O$_3$ NM interface tends to rise higher when β-Ga$_2$O$_3$ NMs SBDs are on a lower-k substrate. The interface temperature of β-Ga$_2$O$_3$ NMs SBDs on the diamond, Si, sapphire, and PI substrate under the biasing condition were calculated to be 298 K, 334 K, 365 K, and 420 K, respectively, which suggests that efficient heat dissipation can occur, or the higher power can be handled when β-Ga$_2$O$_3$ devices are built on a high-k substrate. In addition, the heat distributions of various thick-β-Ga$_2$O$_3$ NMs on different substrates were simulated.

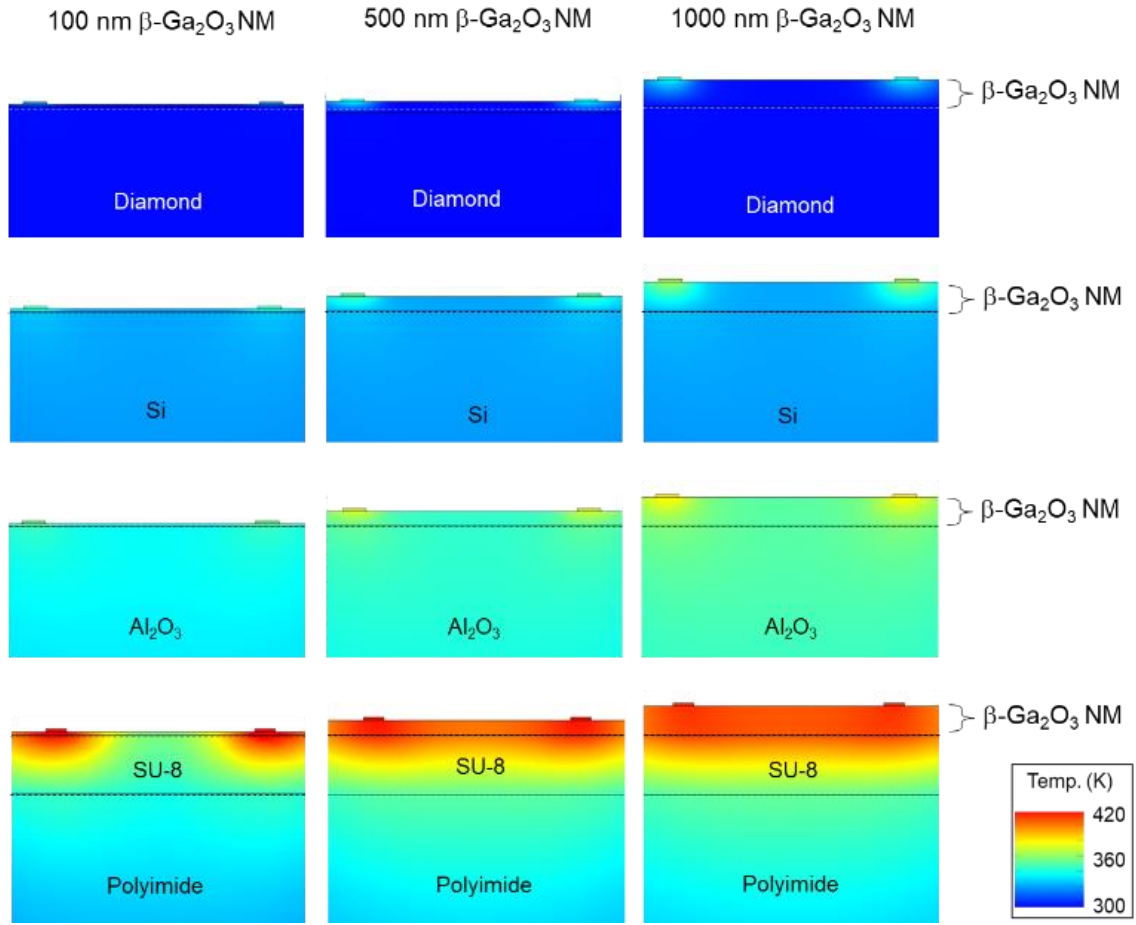

**Figure 5.** Simulated temperature profiles of β-Ga$_2$O$_3$ NMs on four different substrates and three different β-Ga$_2$O$_3$ NM thicknesses.

the empirical material parameters such as density, crystal structure, thermal resistance, and heat capacity from various

Each raw from the left to right of Figure 5 shows the simulation result with 100 nm, 500 nm, and 1000 nm thick β-

**Table 1.** Thermal conductivity values of substrates and β-Ga$_2$O$_3$ used in the simulation.

| | TC |
|---|---|
| PI[35] | 0.25 W/m·K |
| SU-8[36] | 0.2 W/m·K |
| Si[37] | 144 W/m·K |
| Sapphire[38] | 40 W/m·K |
| Diamond[39] | 2200 W/m·K |
| β-Ga$_2$O$_3$ NM[8] | 3.1 W/m·K (100 nm), 5.2 W/m·K (500 nm), 8.68 W/m·K (1000 nm) |

Ga$_2$O$_3$ NM on different substrates. In this simulation, we used the empirical thermal conductivity values of β-Ga$_2$O$_3$ NM by Yixiong et al. and Zheng et al. [8, 27], which shows a decreasing trend in thermal conductivity of β-Ga$_2$O$_3$ NM from 8.7 W/m·K for 1000 nm thick β-Ga$_2$O$_3$ NM to 3.1 W/m·K for 100 nm thick β-Ga$_2$O$_3$ NM. As thermal conductivity becomes smaller as the thickness of β-Ga$_2$O$_3$ NM decreases, one could expect to see a higher temperature in β-Ga$_2$O$_3$ NM as the thickness of β-Ga$_2$O$_3$ NMs decrease. However, the simulation result revealed that the electrode/β-Ga$_2$O$_3$ NM interface temperature decreases as the thickness of β-Ga$_2$O$_3$ NM decreases. While it is true that the thermal conductivity of thinner β-Ga$_2$O$_3$ NM is smaller than the thicker one (i.e., 8.7 W/m·K for 1000 nm thick β-Ga$_2$O$_3$ NM and 3.1 W/m·K for 100 nm thick β-Ga$_2$O$_3$ NM) thus β-Ga$_2$O$_3$ NMs are prone to be heated as β-Ga$_2$O$_3$ NMs get thinner, faster heat dissipation could occur when β-Ga$_2$O$_3$ NMs are thinner and built on a high-k substrate. The simulation result suggests that the devices with a thin β-Ga$_2$O$_3$ NM can dissipate heat faster through a high-k substrate which is critically important to prevent the device from overheating during a biasing condition. However, the opposite trend was observed in the case of β-Ga$_2$O$_3$ NM SBDs on PI substrate which represents a low-k substrate. In other words, the temperature at the electrode/β-Ga$_2$O$_3$ NM interface gets heated several tens of degrees higher to 405 K, 413 K, and 420 K as the thickness of β-Ga$_2$O$_3$ NM is reduced from 1000 nm to 500 nm and 100 nm, respectively. This opposite trend can be explained by the thermal conductivity of β-Ga$_2$O$_3$ NM and PI substrate. As shown in Table 1, the thermal conductivity of β-Ga$_2$O$_3$ NMs (5 W/m·K on average) is much larger than the thermal conductivity of PI and SU-8 (0.25 and 0.2 W/m·K, respectively). Thus, heat dissipation occurs along with the β-Ga$_2$O$_3$ NM, rather than through the PI substrate. Therefore, as the simulation result indicates, the integration of β-Ga$_2$O$_3$ NM with a low-k substrate degrades the contact resistance and affects the sheet resistance and overall bulk properties.

## 3. Conclusions

In conclusion, a transient delayed rising and fall time of β-Ga$_2$O$_3$ NMs SBDs formed on four different substrates were measured using a sub-micron second resolution time-resolved electrical measurement system under a different temperature condition. The devices exhibited noticeably less-delayed turn on/off transient time when β-Ga$_2$O$_3$ NMs SBDs were transfer-printed on a high-k substrate. Furthermore, a relationship between the β-Ga$_2$O$_3$ NM thicknesses and their time-resolved electrical properties was systematically investigated and found that phonon scattering plays an important role in heat dissipation as the thickness of β-Ga$_2$O$_3$ NMs get thinner which is also verified by the Multiphysics simulator. Overall, our result reveals the impact of various substrates with different thermal properties and different β-Ga$_2$O$_3$ NMs thickness with the performance of β-Ga$_2$O$_3$ NMs based devices. Hence, these results can guide further efforts us to optimize the performance of future β-Ga$_2$O$_3$ devices by maximizing heat dissipation from the β-Ga$_2$O$_3$ layer.

## 4. Experimental Section

### 4.1 Device fabrication process

The fabrication process began with a β-Ga$_2$O$_3$ bulk substrate grown by molecular beam epitaxy and moderately Sn doped (concentration of $1\times10^{18}$ cm$^{-3}$) on the [201] direction. The β-Ga$_2$O$_3$ NMs were created by clipping several large segments from the bulk substrate at an angle of 77º, followed by an exfoliation using a well-known taping method. β-Ga$_2$O$_3$ segments were easily mechanically exfoliated in [100] direction due to the weak binding energy in this direction. [23-25, 31] In this step, various thicknesses β-Ga$_2$O$_3$ nanomembranes can be created by adjusting exfoliation times. Once β-Ga$_2$O$_3$ NMs were created, we carefully transfer printed them onto undoped Si, sapphire, diamond, and SU-8 coated polyimide substrates by using an elastomeric stamp (polydimethylsiloxane, PDMS). Prior to the transfer, all the substrates were cleaned thoroughly by the sonification process in acetone, isopropyl alcohol, and DI water for 10 min in each step. After completion of the transfer processes, Ohmic metal and Schottky metal electrodes deposition processes were conducted. Prior to Ohmic metal deposition, a plasma treatment was carried out on β-Ga$_2$O$_3$ NMs by a BCl$_3$/Ar plasma treatment using a reactive ion etcher (RIE) to achieve ohmic contact and to avoid an additional high-temperature annealing process. After that, an Ohmic metal stack (Ti/Au= 20/100 nm) and a Schottky metal stack (Ti/Pt/Au = 20/30/100 nm) were deposited to complete the fabrication of the device.

### 4.2 Measurement setup

The electrical characterization was carried out using a Keithley 4200 semiconductor parameter analyzer. The voltage-current measurement was conducted at voltage bias -4 V to 4 V. To analyze the heat dissipation effect of β-Ga$_2$O$_3$

NMs based diodes, the pulsed I-V setup was implemented as follows: the anode bias is pulsed from off state to on state. The pulse width was 500 μs with a pulse period of 1000 μs. The time resolution of a rise and fall time of pulsed I-V unit was 0.17 μSec. And all devices were measured with the current compliance setting at 3 mA

**Data availability statement**

The data that support the findings of this study are available upon reasonable request from the authors.

**Acknowledgements**

This work was supported by the National Science Foundation (Grant number: ECCS-1809077).

Supplementary Information

for

Transient Characteristics of β-Ga$_2$O$_3$ Nanomembrane Schottky Barrier Diodes on Various Substrates

by

*Junyu Lai, Jung-Hun Seo*[*]*

Department of Materials Design and Innovation,
University at Buffalo, The State University of New York,
Buffalo, NY USA 14260
[*]Email: junghuns@buffalo.edu

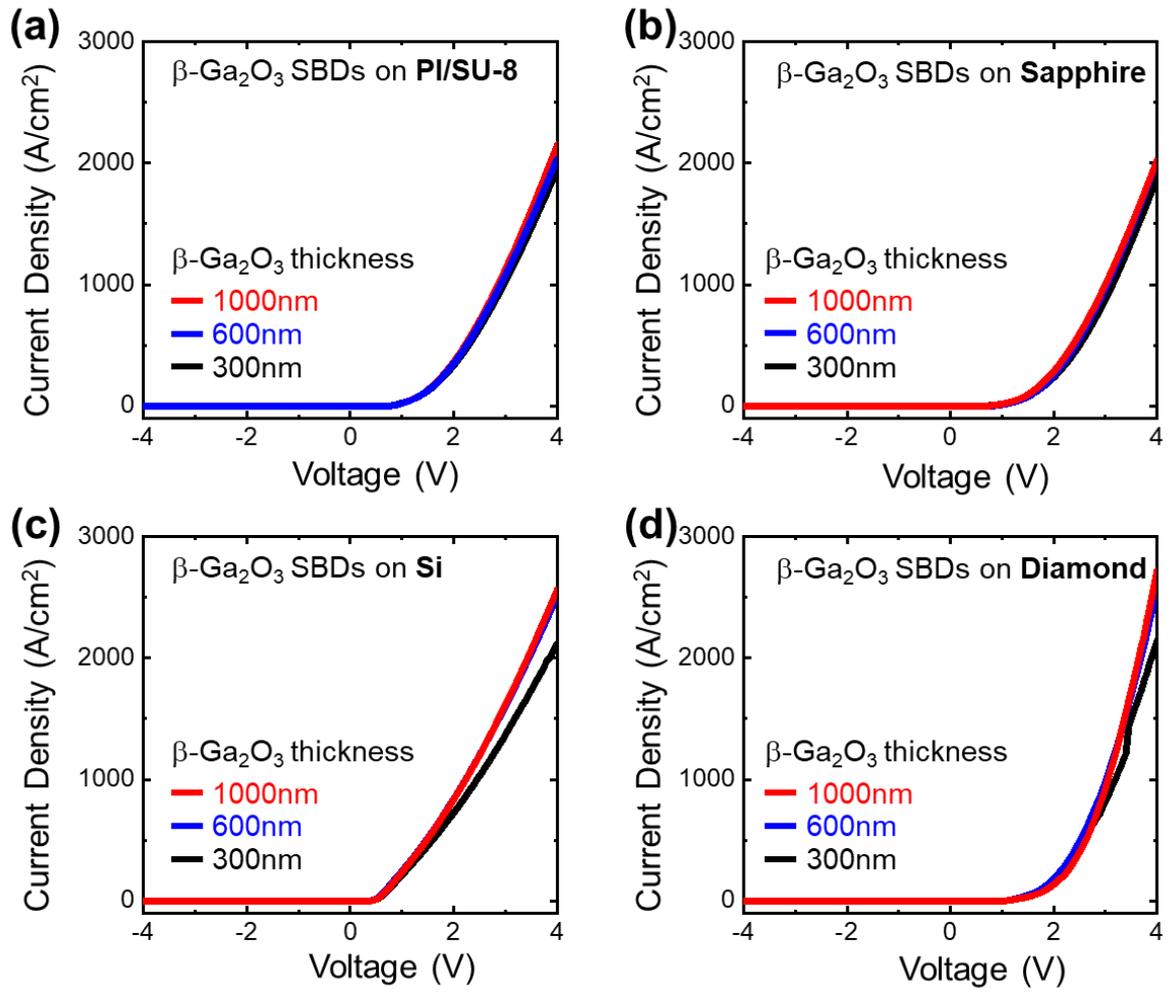

**Figure S1.** I-V characteristics of β-Ga₂O₃ NM SBDs with various NM thicknesses on (a) PI, (b) Si, (c) sapphire, and (d) diamond substrates at -4V to 4V.

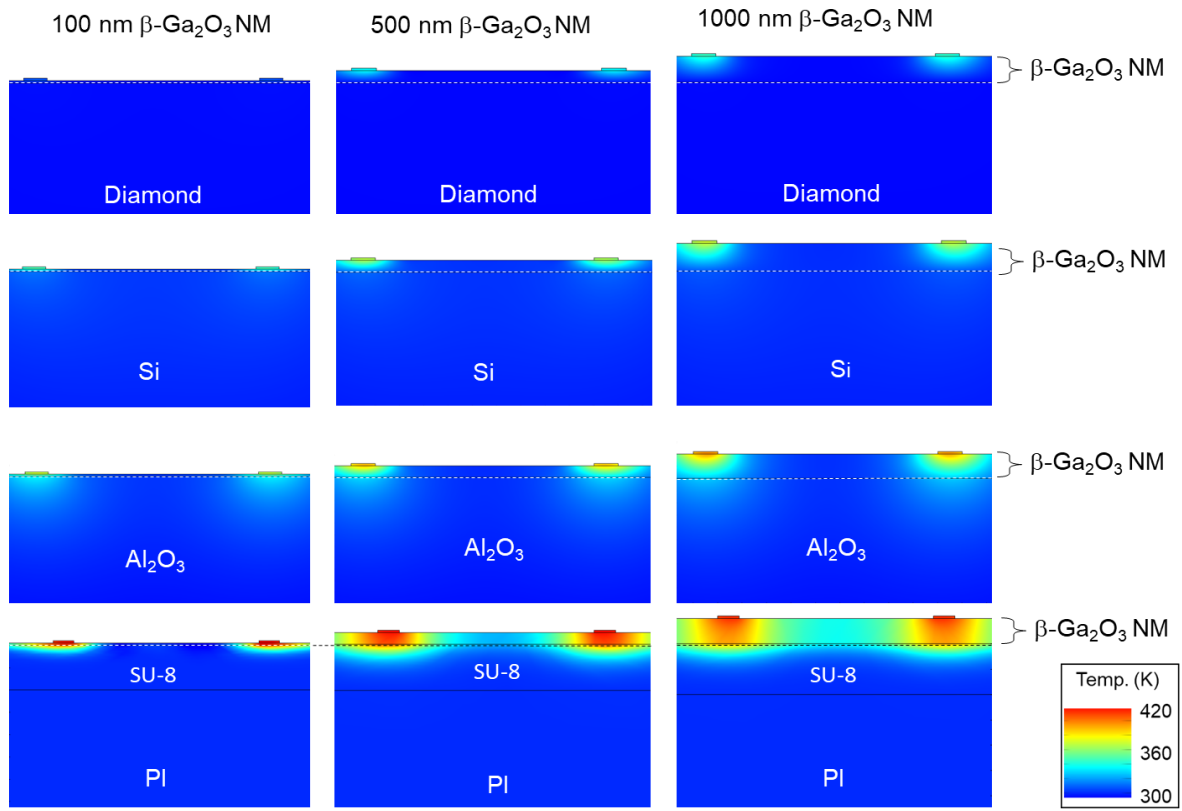

**Figure S2.** Simulated transient temperature profiles (P=0.01W, t =0.01s) of β-Ga$_2$O$_3$ NMs with **(a)** 100 nm, **(b)** 500 nm, and **(c)** 1000 nm thickness, on **(i)** diamond, **(ii)** Si, **(iii)** sapphire, and **(iv)** PI substrates.